\pdfoutput=1
\documentclass[aps,prl,twocolumn,superscriptaddress,showpacs,floatfix,preprintnumbers,nofootinbib]{revtex4-2}
\usepackage[colorlinks,allcolors=blue]{hyperref}
\usepackage{amsmath, amssymb, amsfonts}
\usepackage{graphicx}
\usepackage{bm}
\usepackage{ulem}
\usepackage{color}
\usepackage{xcolor}
\usepackage[mathlines]{lineno}
\usepackage{comment}
\usepackage{multirow}
\usepackage{color}
\usepackage{csquotes}
\usepackage{longtable}
\usepackage{scalerel}
\usepackage[section]{placeins}

\begin{document}
\preprint{TUM-EFT 193/24}
\title{The nature of $\chi_{c1}\left(3872\right)$ and $T_{cc}^+\left(3875\right)$}

\author{Nora Brambilla}
\email{nora.brambilla@tum.de}
\affiliation{Technical University of Munich,\\
TUM School of Natural Sciences, Physics Department,\\   
James-Franck-Str.~1, 85748 Garching, Germany.}
\affiliation{Technical University of Munich, Institute for Advanced Study, \\ 
Lichtenbergstrasse 2 a, 85748 Garching, Germany.}
\affiliation{Technical University of Munich, Munich Data Science Institute, \\ 
Walther-von-Dyck-Strasse 10, 85748 Garching, Germany.}

\author{Abhishek Mohapatra}
\email{abhishek.mohapatra@tum.de}
\affiliation{Technical University of Munich,\\
TUM School of Natural Sciences, Physics Department,\\   
James-Franck-Str.~1, 85748 Garching, Germany.}

\author{Tommaso Scirpa}
\email{tommaso.scirpa@tum.de}
\affiliation{Technical University of Munich,\\
TUM School of Natural Sciences, Physics Department,\\   
James-Franck-Str.~1, 85748 Garching, Germany.}

\author{Antonio Vairo}
\email{antonio.vairo@tum.de}
\affiliation{Technical University of Munich,\\
TUM School of Natural Sciences, Physics Department,\\   
James-Franck-Str.~1, 85748 Garching, Germany.}

\date{\today}

\begin{abstract}
Two decades ago  the $\chi_{c1}\left(3872\right)$ was discovered in the hadron spectrum 
with two heavy quarks. 
The discovery fueled a surge in experimental research, uncovering dozens of so called XYZ exotics states lying outside the conventional quark model, as well 
as theoretical investigations into new forms of matter, such as quark-gluon hybrids, tetraquarks, pentaquarks, with the potential of disclosing new information
about the fundamental strong force.  
Among the XYZs, the $\chi_{c1}\left(3872\right)$ and $T_{cc}^+\left(3875\right)$ stand out for their striking characteristics  and unlashed  many discussions about their nature.
Here, we address this question using  the  Born--Oppenheimer Effective Field Theory (BOEFT) 
and  show  how QCD settles the issue of their composition. 
Not only we describe well the main features of the  $\chi_{c1}\left(3872\right)$ and $T_{cc}^+\left(3875\right)$ but obtain also  predictions in the bottomonium sector.
This opens the way to systematic applications of BOEFT to all XYZs.
\end{abstract}

\maketitle

\paragraph{\textbf{Introduction.}}
The discovery \cite{Belle:2003nnu}  of the $\chi_{c1}(3872)$ (aka X(3872)) state in the charmonium spectrum changed the field of the strong interactions, marking a significant breakthrough. 
A state with $J^{PC} = 1^{++}$ and isospin $I=0$, it is
remarkable for its mass being within 100 keV of the $D^{*0} \bar {D}^0$ threshold.
Considering the central value, it is approximately $50$~keV below the $D^{*0} \bar {D}^0$ threshold and has a very narrow width of $1.19$~MeV \cite{ParticleDataGroup:2024cfk}.
Its properties suggest that it may be a tetraquark, consisting of four quarks bound together, $Q \bar{Q} q \bar{q}$.
Additionally, a plethora of other exotic states, initially termed XYZs, emerged in the sector of the spectrum containing two heavy quarks ($Q\bar{Q}$ or $QQ$) at or above the strong decay thresholds, 
i.e., the energy above which these states can decay into a pair of heavy-light mesons $M\bar{M}$ \cite{ParticleDataGroup:2024cfk,  Brambilla:2019esw}. 
They include states that are electrically charged, like the $Z_c^{ \pm}$, $Z_b^{\pm}$, and pentaquarks $P$  ($Q \bar{Q} q q q$).  
The $T_{cc}^+(3875)$, discovered recently by the LHCb Collaboration \cite{LHCb:2021auc},
stands  out being charged, with a pronounced narrow peak (width of about 0.4 MeV), in the $D^0 D^0 \pi^+$ mass
spectrum, about 360 keV below the $D^{+}D^0$ threshold, with $I=0$ and $J^P=1^+$~\cite{LHCb:2021vvq, LHCb:2021auc}.
It is directly produced in hadroproduction and is the longest-lived exotic matter particle ever found.
The XYZ states have been studied  at the existing collider experiments and will be studied 
at the upcoming new experiments at FAIR~\cite{PANDA:2021ozp} and EIC~\cite{Burkert:2022hjz}. 
The $\chi_{c1}(3872)$ has been observed with  large ($\sim  30$~nb) production cross sections both at CDF~\cite{CDF:2006ocq} and CMS~\cite{CMS:2013fpt} and in heavy ion collisions~\cite{CMS:2021znk}.  
It is observed also in $e^+ e^-$ production and in $B$ decays. 
Its decays to charm hidden states, some decay branching ratios, the production in $e^+ e^- \to \omega \chi_{c1}\left(3872\right)$ \cite{BESIII:2022bse},  and the  compositeness value   \cite{BESIII:2023hml, Esposito:2021vhu} point to  a charmonium component in its Fock space decomposition.
The characteristics of these exotics reveal them as novel, strongly correlated systems with the potential of disclosing information about the strong force.

The existence of exotic hadron states with more than the minimal quark content  was already proposed  in the quark model  \cite{Gell-Mann:1964ewy,Zweig:1964ruk,Gross:2022hyw} and 
in QCD \cite{Fritzsch:1973pi,Gross:2022hyw} with the addition of states involving  gluons in the binding, the hybrids.
The search for exotic states started in the sixties, but firm evidence came only when the XYZs were discovered.
These new structures  triggered a huge theoretical effort.  
The simplest system consisting of two quarks and two antiquarks (tetraquark) is already a very complicated one 
and it is unclear whether any kind of clustering occurs in it. 
To simplify the problem, models have been based on ad hoc choices of dominant configurations and interactions
\cite{Brambilla:2022ura,Brambilla:2019esw,Brambilla:2014jmp,Brambilla:2010cs,Guo:2017jvc,Karliner:2017qhf,Ali:2017jda,Bodwin:2013nua,Maiani:2020pur,Ali:2019roi,Godfrey:2008nc,Lebed:2016hpi,Mezzadri:2022loq,Gross:2022hyw,Olsen:2017bmm}. For the tetraquark examples  include hadronic molecules, which assume constituent color singlet mesons bound by residual nuclear forces  \cite{Tornqvist:2004qy,Close:2003sg,Guo:2017jvc}, and compact tetraquarks, which assume bound states between colored diquarks and antidiquarks~\cite{Jaffe:1976ig,Jaffe:2003sg,Esposito:2016noz,Maiani:2004vq}.
For a long time, these two modelizations competed in asserting the nature of the XYZs,  
and the $\chi_{c1}(3872)$ in particular~\cite{Esposito:2021vhu,Baru:2021ldu}.

QCD lattice calculations of some  XYZ masses exist \cite{HadronSpectrum:2012gic, Cheung:2016bym,Ryan:2020iog,  Brown:2012tm, Prelovsek:2013cra, Padmanath:2015era,Bicudo:2012qt,Bicudo:2015kna,Bicudo:2015vta,Bicudo:2016ooe,Leskovec:2019ioa,Meinel:2022lzo,Alexandrou:2023cqg,Alexandrou:2024iwi, Aoki:2023nzp,Mohanta:2020eed,Junnarkar:2018twb,Francis:2016hui,Francis:2018jyb,Hudspith:2020tdf,Padmanath:2022cvl,Hudspith:2023loy,Colquhoun:2024jzh, Padmanath:2023rdu, Radhakrishnan:2024ihu} but remain challenging as they require studying coupled channel scatterings on the lattice~\cite{Luscher:1990ux,Briceno:2017max}.

\paragraph{\textbf{Born--Oppenheimer EFT.}}
BOEFT is a nonrelativistic  effective field theory obtained from QCD on the basis of symmetry and scales separation. 
It is suitable to describe bound states of two heavy quarks and any light degrees of freedom (LDF).
Strongly coupled potential nonrelativistic QCD (pNRQCD) for quarkonium is the simplest realization of this theory \cite{Brambilla:1999xf,Brambilla:2000gk,Pineda:2000sz,Brambilla:2002nu}. 
The original BOEFT for hybrids was obtained in~\cite{Berwein:2015vca}, with subsequent work~\cite{Brambilla:2017uyf, Oncala:2017hop, Soto:2020xpm, Braaten:2024tbm} up to~\cite{Berwein:2024ztx}, 
where the equations for tetraquarks and pentaquarks were derived. 

The first scale separation exploited by BOEFT is the scale separation between the mass $m_Q$ of the heavy quarks and the energy scale of the LDF, gluons or light quarks part of the binding, which is of the order of the nonperturbative hadronic scale $\Lambda_{\rm QCD}$: $m_Q \gg \Lambda_{\rm QCD}$. 
QCD static energies for all LDF excitations in the presence of two quarks separated by a distance $r$ are classified in terms of the {\it BO quantum numbers}. 
Given ${\bm K}$, the total LDF  angular momentum, we define $\Lambda =|\hat{r}\cdot {\bm K}|$ ($\Lambda = 0, 1, 2, \dots$ are denoted by $\Sigma, \Pi, \Delta, \dots$).  
The BO quantum numbers $\Lambda^{\sigma}_{\eta}$ identify the cylindrical $D_{\infty h}$ representations, where $\eta=g, u$ is the $[C]P$ eigenvalue $\pm 1$ (for $QQ$ only parity is defined), and  $\sigma=\pm1$ (only for $\Sigma)$ is the eigenvalue of reflection through a plane containing the two sources. 
Excited states  are indicated as  $\Lambda_\eta^{\sigma\prime}$, $\Lambda_\eta^{\sigma\prime\prime}$, \ldots.
We may associate to the LDF also the label $\kappa\equiv\{k^{P[C]},f\}$, 
where $k\left(k+1\right)$ is the eigenvalue of ${\bm K}^2$ and $f$ is the flavor. 
As $r\rightarrow 0$, the cylindrical symmetry reduces to a spherical symmetry labeled by $\kappa$, and some BO static energies become degenerate.
The QCD static energies are written in terms of gauge invariant generalized Wilson loops characterized by the BO quantum numbers~\cite{Berwein:2024ztx}. 
Some of these Wilson loops have been computed in lattice QCD~\cite{Juge:1999ie,Juge:2002br,Bali:2000vr,Bali:2003jq,Capitani:2018rox,Schlosser:2021wnr, Bicudo:2021tsc, Sharifian:2023idc, Hollwieser:2023bud, Brown:2012tm, Bicudo:2015kna, Prelovsek:2019ywc,Sadl:2021bme,Mueller:2023wzd}.

The next scale separation is the one between the energy of the LDF and the binding energy 
of the heavy quarks: $ \Lambda_{\rm QCD} \gg E$. 
This scale separation allows to define the BOEFT and obtain coupled Schr\"odinger equations to describe the exotics.  
The potentials appearing in these equations are the static energies introduced above.
The potentials mix.
The mixing may occur either because the potentials become degenerate as $r\rightarrow 0$ or because they carry the same BO quantum numbers and get close (degenerate) at some finite distance~\cite{Berwein:2024ztx}.

We present first-time predictions for the $\chi_{c1}(3872)$ obtained from Schrödinger coupled equations derived from the BOEFT, and results for the $T_{cc}^+(3875)$ obtained from a  single-channel equation, whose $T_{bb}$ analog has been studied previously in \cite{Brown:2012tm, Bicudo:2012qt,Bicudo:2015kna,Bicudo:2015vta, Bicudo:2016ooe}.
Moreover, in \cite{Berwein:2024ztx}, we have shown that BOEFT constrains the potentials at short and large distances.
At short distances $r$, we rely on the multipole expansion and express the potentials as a power series in $r$.
The constant term in the series is the adjoint meson mass in the case of $\chi_{c1}(3872)$ 
and the triplet meson mass in the case of $T_{cc}^+(3875)$; it depends only on $\kappa$. 
At large distances, due to BO quantum numbers conservation, the potentials evolve into the $M\bar{M}$ static energies. 
The adjoint or triplet meson masses are the key parameters for the existence of states close to the heavy-light meson thresholds \cite{Braaten:2024tbm}.

\paragraph{\textbf{$\chi_{c1}\left(3872\right)$: Coupled Schr\"odinger equations and potentials.}}
In the BO picture, we take the state $\chi_{c1}(3872)$  as a bound state below the spin-isospin averaged $D\bar{D}$ threshold of the lowest $Q\bar{Q}$ tetraquark potential associated with the isospin-0 $k^{PC}=1^{--}$ adjoint meson.  
Without spin interactions, different spin combinations, $\{S=0, S=1\}$ with $S$ the heavy-quark-pair spin,  
form degenerate multiplets. 
So, the ground-state $J^{PC}$ multiplet for  $k^{PC}=0^{-+}$ is $\{0^{++}, 1^{+-}\}$  and  for $k^{PC}=1^{--}$ is $\{1^{+-}, \left(0, 1, 2\right)^{++}\}$. 
We identify $\chi_{c1}(3872)$ with the $1^{++}$ state in the ground-state multiplet 
corresponding to $k^{PC} = 1^{--}$ and described by the BO potentials $V_{\Sigma_g^{+\prime}}$ and $V_{\Pi_g}$.

The potentials $V_{\Sigma_g^{+\prime}}$ and $V_{\Pi_g}$ mix at short distance and with the
quarkonium $V_{\Sigma_g^+}$ potential due to avoided level crossing at around $1.2$~fm, see Fig.~\ref{fig:isospin0},
giving rise to coupled channel Schr\"odinger equations (Eq.~\eqref{eq:Sch-QQbar} in  Supplemental Material) \cite{Berwein:2024ztx}.
The equations depend on the angular momentum $\bm{L}=\bm{L}_{Q}+\bm{K}$, with $\bm{L}_{Q}$ the heavy quark pair orbital angular momentum; $l(l+1)$ are the eigenvalues of $\bm{L}^2$.

We use the lattice parametrization of Ref.~\cite{Bulava:2024jpj} 
to model $V_{\Sigma_g^+}$ and $V_{\Sigma_g^{+\prime}}$ around the string breaking region.
For $V_{\Sigma_g^{+\prime}}$ and $V_{\Pi_g}$,  we model the short-distance behavior using the quenched BO-potential parametrization from \cite{Alasiri:2024nue}, due to lack of lattice data, and the long-distance behavior with a two-pion exchange potential \cite{Lyu:2023xro}:
\begin{align}
&V_{\Sigma_g^+}(r) = V_0 + \frac{\gamma}{r} + \sigma r,\label{eq:Vcornell}\\
&V_{\Lambda}(r) =
   \begin{cases}
\kappa_8/r +  \Lambda_{1^{--}} + A_{\Lambda}\, r^2+ B_{\Lambda} \, r^4
\; r < R_{\Lambda}
\\
F_{\Lambda}\,e^{-r/d}/r^2 + E_1 
\quad\quad\qquad\,\,\,\,\,\,  ~~ r > R_{\Lambda}
\end{cases}\!\!\!\!\!\!, 
\label{eq:QQbar_V}
\end{align}
where $\Lambda\in\{\Sigma_g^{+\prime}, \Pi_g\}$, $\gamma = -0.434$, $\sigma = 0.198\,\mathrm{GeV}^2$, $\kappa_8=0.037$, $A_{\Sigma_g^{+\prime}} = 0.0065 \,\mathrm{GeV}^3$, $B_{\Sigma_g^{+\prime}} = 0.0018\,\mathrm{GeV}^5$, $A_{\Pi_g} = 0.0726\,\mathrm{GeV}^3$, $B_{\Pi_g}=-0.0051\,\mathrm{GeV}^5$, $d \sim 1/\left(2m_\pi\right) \sim 1/0.3\,\mathrm{GeV}^{-1}\sim 0.66\,\mathrm{fm}$, $E_1=0.005$~GeV, $V_0 = -1.142$~GeV, and the parameters $F_{\Lambda}$ and $R_{\Lambda}$ are determined by imposing continuity up to first derivatives. 
The mixing potential $V_{\Sigma_g^+ -\Sigma_g^{+\prime}}$ must vanish at $r\rightarrow 0$ based on pNRQCD~\cite{TarrusCastella:2022rxb} and it approaches zero asymptotically at $r \to \infty$, with a peak near the string-breaking region. 
Hence, we parameterize $V_{\Sigma_g^+ -\Sigma_g^{+\prime}}$ as
\begin{align}
  V_{\Sigma_g^+ -\Sigma_g^{+\prime}} = 
  \begin{cases}
g\,r/r_1 \quad &r < r_1
\\
g\, 
\quad  &r_1\le r \le r_2
\\
g \, e^{-(r-r_2)/r_0} & r > r_2
\end{cases}, 
\label{eq:Vmix}
\end{align}
where $r_0=0.5$~fm is the Sommer scale, 
and $g=0.05$~GeV, $r_1=0.95$~fm and $r_2=1.51$~fm are fixed on the lattice data \cite{Bulava:2024jpj} and by demanding the continuity of the potential. 

\begin{figure}[ht]
\centerline{ \includegraphics*[width=7.5cm,clip=true]{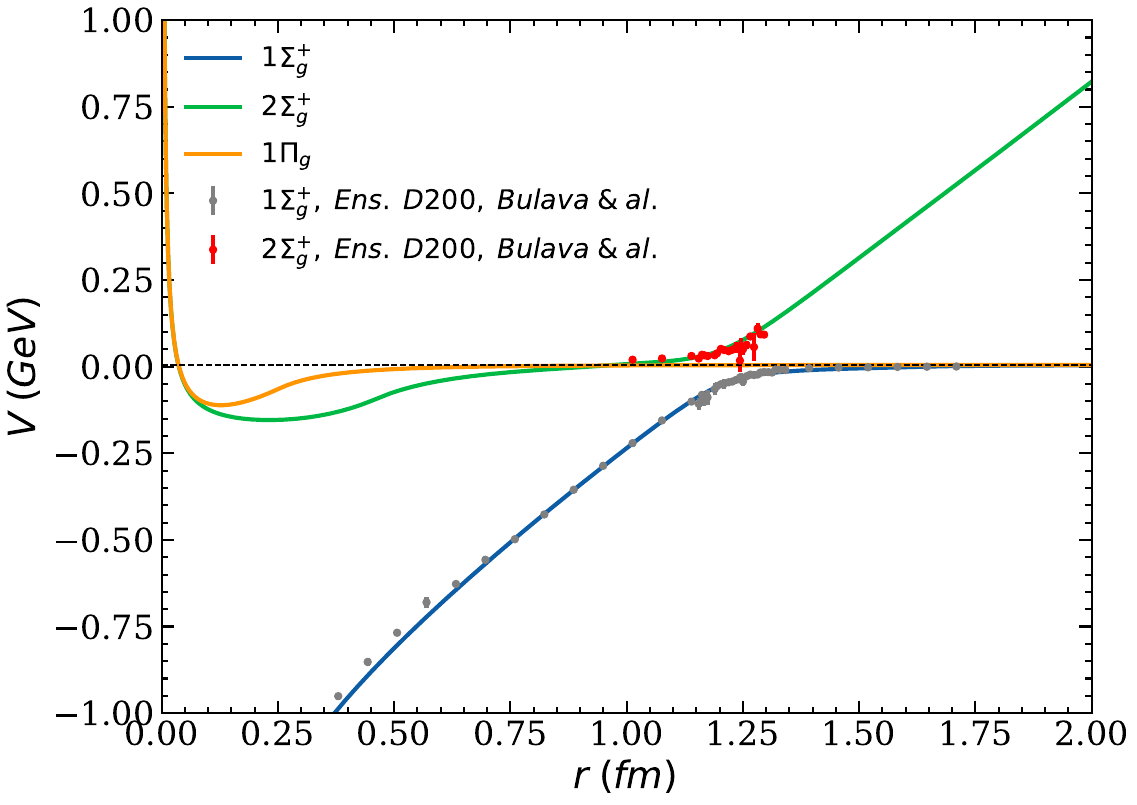}}
\caption{Lowest isospin-0 adiabatic $Q\bar{Q}$ tetraquark BO potentials as functions of $r$~\cite{Berwein:2024ztx}, with $V_{2\Sigma_g^{+}}$ and $V_{1\Pi_g}$, corresponding to the $1^{--}$ adjoint meson at short distances (note the repulsive behavior of the color octet potential), approaching from below the $M\bar{M}$ threshold set at $0.005$~GeV. The {\it adiabatic potentials} are the eigenvalues of the potential matrix defined in Eqs.~\eqref{eq:Vcornell}-\eqref{eq:Vmix} (see Eqs.~\eqref{eq:V1S}-\eqref{eq:V1P}) \cite{Berwein:2024ztx}. We show our potentials  along with lattice data from the D200 ensemble \cite{Bulava:2024jpj} with $2\Sigma_g^{+}$ data limited to the avoided crossing region.}
\label{fig:isospin0}
\end{figure}

The charm and bottom quark masses entering the kinetic energy in the Schr\"odinger
equations are set to be the spin-isospin averaged $D$ and $B$ meson masses, $m_c=1.973$~GeV and $m_b=5.313$~GeV.
The physical spin-isospin averaged meson-(antimeson) thresholds that we use in this work 
are $E_{D\bar{D}}=E_{DD}=3.946$~GeV, $E_{B\bar{B}}=E_{BB}=10.627$~GeV, and 
$E_{DB}=7.287$~GeV~\cite{ParticleDataGroup:2024cfk}.
We treat the $1^{--}$ adjoint meson mass $\Lambda_{1^{--}}$ as a free parameter to be fixed 
on the $\chi_{c1}\left(3872\right)$ mass.

\paragraph{\textbf{Results for $\chi_{c1}\left(3872\right)$ and $\chi_{b1}$.}}
The  binding energy depends on the  $1^{--}$ adjoint meson mass $\Lambda_{1^{--}}$. 
Solving the coupled Schr\"odinger equations with the potentials in Eqs.~\eqref{eq:Vcornell}-\eqref{eq:Vmix} and $\Lambda_{1^{--}}$ as a free parameter, 
we obtain a bound state around $90$~keV below the $D\bar{D}$ threshold, which we identify with $\chi_{c1}\left(3872\right)$, for the value $\Lambda_{1^{--}}^* = 919$~ MeV. 
The scattering length $a$ corresponding to the $E_b = 90$~keV binding energy is $a=\hbar/\sqrt{m_c E_b}= 14.81$~fm. 
Additionally, we find a deeper bound state around $409$~MeV below the $D\bar{D}$ threshold with mass  $3537$~MeV, 
which we identify with the spin-averaged $\chi_{c}\left(1P\right)$ state.

Our adjoint meson mass $\Lambda_{1^{--}}^*$ is 
consistent with the estimate from quenched lattice QCD with valence quarks \cite{Foster:1998wu}
and changes only  within $50$ to $80$~MeV depending on the one pion/two pions long-distance parameterization of $V_{\Sigma_g^{+\prime}}$ and $V_{\Pi_g}$.
The  $\chi_{c1}(3872)$ composition entails $8 \%$ of quarkonium  and $92\%$
of tetraquark, split into  $P_{\Sigma^{\prime}}=38 \%$ and $P_{\Pi}=54 \% $.
These percentages show very little dependence 
on the specific parametrization of the long-distance part of the mixing potential in Eq.~\eqref{eq:Vmix}, e.g., whether it approaches zero with an exponential, power law, or Gaussian profile. 
Instead, increasing $g$ up to $0.06$~GeV and adjusting  $\Lambda_{1^{--}}^*$ within $60$~MeV to get again the $\chi_{c1}(3872)$ raises the quarkonium fraction to $13 \%$. Notably, for a state exactly at the threshold, the tetraquark probability dominates, $P_{\Sigma^{\prime}}+P_{\Pi}\gtrsim 99\%$, with a negligible 
quarkonium probability, $P_{\Sigma}\lesssim 1\%$;
the opposite happens for a deeper state like $\chi_{c}\left(1P\right)$.

The radiative decays of $\chi_{c1}\left(3872\right)$, $\chi_{c1}\left(3872\right)\rightarrow \gamma J/\psi$ and $\chi_{c1}\left(3872\right)\rightarrow \gamma\psi\left(2s\right)$,  provide key insights into its internal structure and whether it has a significant charmonium component. 
In particular, the ratio 
${\cal R}_{\gamma\psi}={\Gamma_{\chi_{c1}\left(3872\right)\rightarrow \gamma\psi\left(2s\right)}}/{\Gamma_{\chi_{c1}\left(3872\right)\rightarrow \gamma J/\psi}}$, 
has been subject of  significant theoretical \cite{Swanson:2003tb,Swanson:2004pp, Dong:2009uf, Guo:2014taa,
Molnar:2016dbo,Rathaud:2016tys, Lebed:2022vks, Grinstein:2024rcu, Barnes:2003vb, Barnes:2005pb,DeFazio:2008xq,Li:2009zu, Dong:2009uf, Badalian:2012jz, Ferretti:2014xqa, Badalian:2015dha, Deng:2016stx, Giacosa:2019zxw} and experimental \cite{BaBar:2008flx, Belle:2011wdj, LHCb:2014jvf, BESIII:2020nbj, LHCb:2024tpv} interest in deciphering the nature of $\chi_{c1}\left(3872\right)$. 
Theoretical predictions for $R_{\gamma\psi}$ from different models are given in Table 1 of Ref.~\cite{LHCb:2024tpv}: 
$R_{\gamma\psi}\ll 1$ is expected for a pure $D\bar{D}^*$ molecule~\cite{Swanson:2003tb,Swanson:2004pp, Dong:2009uf, Rathaud:2016tys, Lebed:2022vks, Grinstein:2024rcu}, while $R_{\gamma\psi}\gtrsim 1$ is expected for 
a $\chi_{c1}\left(2P\right)$ charmonium-molecule mixture or a pure $\chi_{c1}\left(2P\right)$ charmonium state 
\cite{Barnes:2003vb, Barnes:2005pb,DeFazio:2008xq,Li:2009zu, Dong:2009uf, Badalian:2012jz, Ferretti:2014xqa, Badalian:2015dha, Deng:2016stx, Giacosa:2019zxw}. 
Recently, based on Run 1 and Run 2 data, the LHCb collaboration reported the updated value $R_{\gamma\psi}=1.67 \pm 0.25$ \cite{LHCb:2024tpv}. 
As mentioned above, we obtain naturally a $8\%$ $\chi_{c1}\left(2P\right)$ charmonium component in $\chi_{c1}\left(3872\right)$ from the mixing due to the avoided level crossing. 
Assuming the radiative decay is going only through quarkonium component and considering  the average of the several quark model results for the $\chi_{c1}\left(2P\right)\rightarrow \gamma J/\psi$ and $\chi_{c1}\left(2P\right)\rightarrow \gamma\psi\left(2s\right)$ widths given in Table 1 of Ref.~\cite{Pei:2024hzv},
our rough estimates for $\chi_{c1}\left(3872\right)$, based on a $8\%$ charmonium component, are $\Gamma_{\chi_{c1}\left(3872\right)\rightarrow \gamma J/\psi}=3.9 \pm 2.6$~keV and $\Gamma_{\chi_{c1}\left(3872\right)\rightarrow \gamma\psi\left(2s\right)}=11.7 \pm 5.1$~keV, which are consistent with the experimental results  $\Gamma^{\mathrm{PDG}}_{\chi_{c1}\left(3872\right)\rightarrow \gamma J/\psi}=9.3 \pm 5.1$~keV and $\Gamma^{\mathrm{PDG}}_{\chi_{c1}\left(3872\right)\rightarrow \gamma \psi\left(2s\right)}=54 \pm 33$~keV \cite{ParticleDataGroup:2024cfk}. 
Our prediction for the ratio is $R_{\gamma \psi}=2.99 \pm 2.36$, 
which agrees with the recent LHCb value within errors. 
Moreover, our result indicating a $c\bar{c}$ component in $\chi_{c1}\left(3872\right)$ between 8\% and 13\%  is close to the $5\%$ estimate mentioned in some literature to explain the production rate of $\chi_{c1}\left(3872\right)$ in $pp$ collisions~\cite{Bignamini:2009sk, Takizawa:2012hy, ATLAS:2016kwu, BESIII:2022bse}.
Additionally, using Weinberg compositeness criterion \cite{Weinberg:1962hj, Weinberg:1965zz}, which defines $Z$ as the probability of finding a compact component in the bound state,  
the BESIII collaboration reported $Z=0.18^{+0.20}_{-0.23}$ based on the coupled channel analysis of the $\chi_{c1}\left(3872\right)$ lineshape~\cite{BESIII:2023hml}.
Ref.~\cite{Esposito:2021vhu} reported $0.052<Z<0.14$ based on an analysis of high-statistics LHCb data for the $\chi_{c1}\left(3872\right)$ lineshape~\cite{LHCb:2020xds}.   
Within error bars, these values for $Z$ are consistent with our 
estimate of the $c\bar{c}$ component in $\chi_{c1}\left(3872\right)$.
Finally, also the lattice studies in Refs.~\cite{Prelovsek:2013cra, Padmanath:2015era} suggest the existence of a quarkonium component.
With  $\Lambda_{1^{--}}^*$ tuned on the $\chi_{c1}\left(3872\right)$ mass and the potentials in Eqs.~\eqref{eq:Vcornell}-\eqref{eq:Vmix}, we predict in the bottomonium sector 
a bound state about $15$~MeV below the $B\bar{B}$ threshold, with quarkonium probability around $1.5\%$.
Additionally, we find three deeper bound states around $744$~MeV, $391$~MeV, and $108$~MeV below the $B\bar{B}$ threshold with masses $9.883$~GeV, $10.236$~GeV, and $10.519$~GeV respectively, 
which we identify with the spin-averaged $\chi_{b}\left(1P\right)$, $\chi_{b}\left(2P\right)$, and $\chi_{b}\left(3P\right)$ states.

\paragraph{\textbf{Spin splitting.}}
Considering the spin-dependent  corrections to the potentials, we can determine the spin-splittings. 
For the $k^{PC}=1^{--}$ adjoint meson, the lowest $Q\bar{Q}$ tetraquark multiplet (with $l=1$)
includes a $S=0$ state $1^{+-}$ and three  $S=1$ states $(0,1,2)^{++}$.
The BOEFT predicts that spin-dependent interactions arise already at order $1/m_Q$ in tetraquarks and in hybrids, 
differently from quarkonium \cite{Oncala:2017hop, Brambilla:2018pyn, Brambilla:2019jfi, Soto:2023lbh}.

\begin{figure}[ht]
\centerline{ \includegraphics*[width=8.0cm,clip=true]{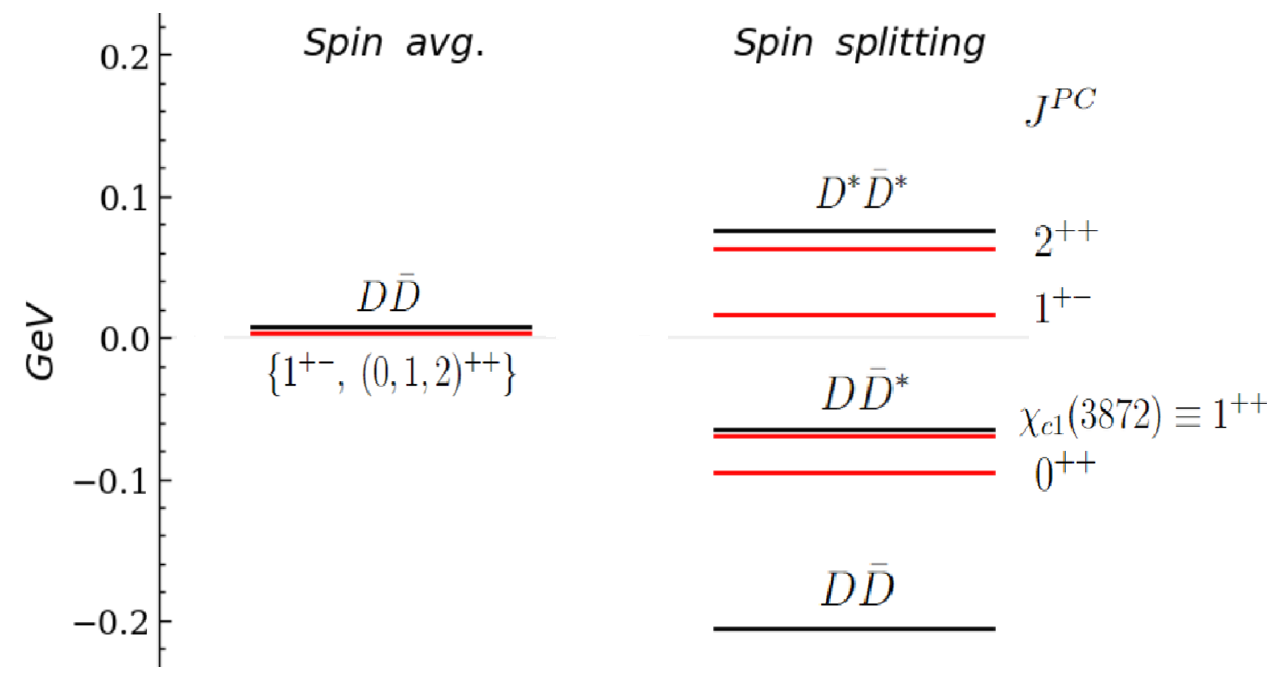}}
\caption{Spectrum of the lowest $c\bar{c}$ tetraquark multiplet $\{1^{+-}, \left(0, 1, 2\right)^{++}\}$ corresponding to the $1^{--}$ adjoint meson relative to the heavy meson pair threhsold. On the left, we display the spin-averaged case, and on the right, we display the spectrum with respect to the $D\bar{D}$, $D^*\bar{D}$ (or $D\bar{D}^*$) and $D^*\bar{D}^*$ thresholds after including spin-corrections.}
\label{fig:spin1--}
\end{figure}
The spin-dependent potentials depend on nonperturbative correlators that for hybrids have been fixed~\cite{Brambilla:2018pyn, Brambilla:2019jfi} on a direct lattice calculation of the charmonium hybrid masses~\cite{HadronSpectrum:2012gic, Cheung:2016bym}.
Currently, no lattice calculations for the spin splittings of the tetraquark multiplets exist. 
Therefore, we try to estimate the spin corrections for the lowest $c\bar{c}$ tetraquark  by averaging the lattice results from \cite{HadronSpectrum:2012gic, Cheung:2016bym} for the spin structure of the lowest $c\bar{c}$ hybrid multiplet. 
We also account for the  spin splitting $\delta_Q$, which scales like $1/m_Q$, between heavy mesons $M^*$ and $M$: 
$\delta_c=141$~MeV and $\delta_b=45$~MeV~\cite{ParticleDataGroup:2024cfk}. 
The energies of $M$ and $M^*$ relative to their spin average are $-3\delta_Q/4$ and $\delta_Q/4$, respectively. 
This implies that the isospin averaged thresholds $D\bar{D}$ and  $D^*\bar D$ (or  $D \bar D^*$) are about $212$~MeV and $70.5$~MeV below, 
while $D^*\bar D^*$ is about $70.5$~MeV above the spin-isospin averaged value. 
With the spin corrections from \cite{HadronSpectrum:2012gic, Cheung:2016bym}, the  $1^{+-}$ and $2^{++}$ states are shifted above by $11\left(11\right)$~MeV and $58\left(14\right)$~MeV, 
while the $0^{++}$ and $1^{++}$ states are shifted below by $100\left(11\right)$~MeV and $74\left(14\right)$~MeV 
relative to the spin averaged (see Fig.~\ref{fig:spin1--}). The values in parentheses represent the lattice statistical errors of \cite{HadronSpectrum:2012gic, Cheung:2016bym} combined in quadrature. An alternative estimate~\cite{Braaten:2024tbm} 
of these corrections due to $M\bar{M}$ threshold splittings is given in the Supplemental Material and agrees inside errors.
The $\chi_{c1}\left(3872\right)$, identified with the $1^{++}$ state, is about  $3.5\left(14\right)$~MeV below the isospin averaged $D^*\bar{D}$ threshold, 
which is consistent with the experimental fact that the state is $100$~keV within $D^{*0} \bar {D}^0$ and $4$~MeV below the isospin averaged $D^*\bar{D}$ threshold \cite{ParticleDataGroup:2024cfk}. 
Given $E_{D\bar{D}}=3.946$~GeV, the masses of the $1^{+-}$, $0^{++}$, $1^{++}$, and $2^{++}$ states are $3.957\left(11\right)$~GeV,  $3.846\left(11\right)$~GeV,  $3.872\left(14\right)$~GeV, and $4.004\left(14\right)$~GeV, respectively.
Moreover, the $1^{+-}$ state could be identified with $X\left(3940\right)$ given that the masses are consistent within errors~\cite{Belle:2007woe}. Spin splittings in the bottom sector are discussed in the  Supplemental Material.

\begin{figure}[ht]
\centerline{ \includegraphics*[width=7.5cm,clip=true]{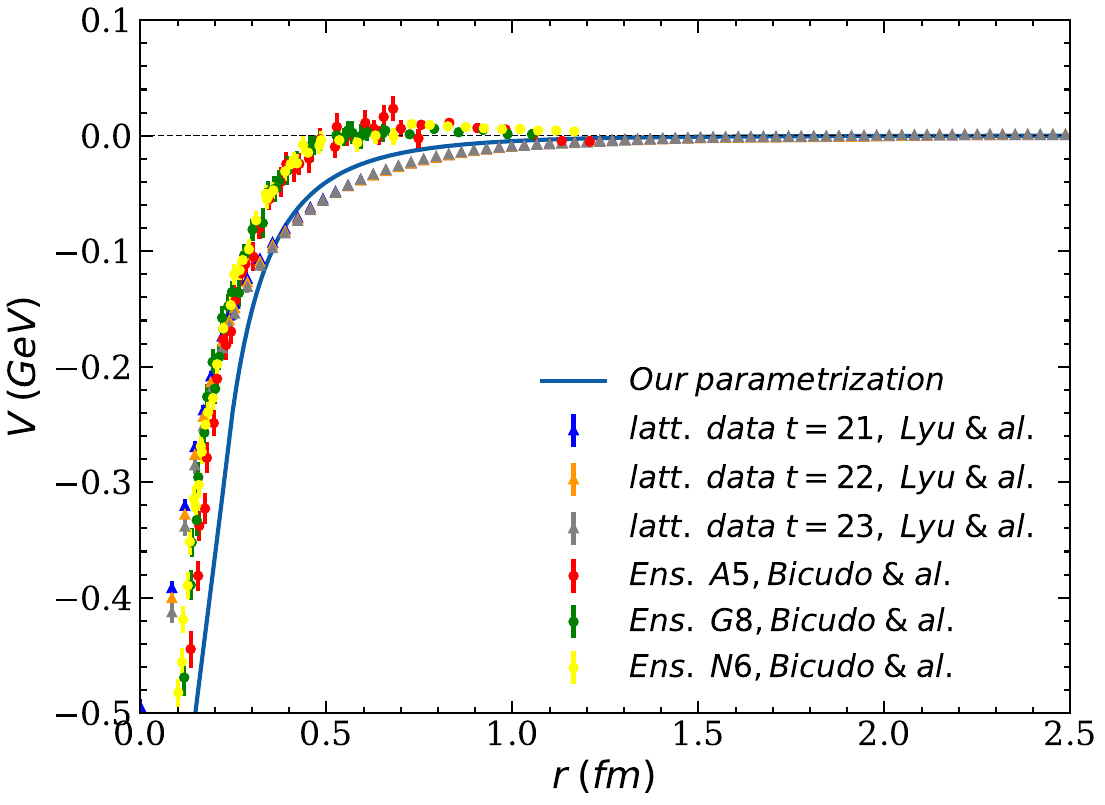}}
\caption{ Lowest isospin-0 $QQ$ tetraquark BO potential as a function of $r$~\cite{Berwein:2024ztx}.The BO potential $V_{\Sigma_g^{+}}$, corresponding to the $0^{+}$ triplet meson at $r\to 0$, approaches at $r\to\infty$ the $MM$ threshold, which is set to zero energy. Our potential is depicted alongside the lattice data from~\cite{Bicudo:2024vxq,Lyu:2023xro}.}
\label{fig:isospin1}
\end{figure}

\paragraph{\textbf{$T_{cc}^+\left(3875\right)$: Schr\"odinger equations and potential.}}
In the BOEFT, we take the state $T_{cc}^+(3875)$ as a bound state below the spin-isospin averaged $DD$ threshold of the lowest $QQ$ tetraquark potential associated with the isospin-0 $k^{P}=0^{+}$ triplet meson. 
For  $k^{P}=0^{+}$, due to the Pauli exclusion principle, the ground-state multiplet only includes a heavy-quark-pair spin-triplet $\left(S=1\right)$ state $1^{+}$~\cite{Berwein:2024ztx}.   
We identify $T_{cc}^+(3875)$ with this state in the ground-state multiplet (with $l=0$)
of the BO potential $V_{\Sigma_g^{+}}$ associated at $r\to 0$ with the $0^{+}$ triplet meson. 
In this case, we have a single channel Schr\"odinger equation to solve (Eq.~\eqref{eq:SchQQ} in Supplemental Material),
with the potential  $V_{\Sigma_{g}^+}$ based on the short-distance behavior of~\cite{Alasiri:2024nue} and the long-distance behavior induced by a two-pion exchange potential~\cite{Lyu:2023xro}: 
\begin{align}
V_{\Sigma_{g}^+}=
\begin{cases}
\kappa_3/r +  \Lambda_{0^{+}} + A_{\Sigma_g^+}\, r^2
\quad & r < R_{\Sigma_g^+}
\\
F_{\Sigma_g^+}\,e^{-r/d}/r^2 
\quad & r > R_{\Sigma_g^+}
\end{cases},
\label{eq:VQQ}
\end{align}
where $\kappa_3=-0.120$ and $A_{\Sigma_g^+} = 0.197 \,\mathrm{GeV}^3$ \cite{Alasiri:2024nue}, 
while the parameters $F_{\Sigma_g^+}$ and $R_{\Sigma_g^+}$ are determined by imposing continuity up to the first derivatives. 
In Fig.~\ref{fig:isospin1}, we compare our parametrization of the potential  $V_{\Sigma_{g}^{+}}(r)$ for isospin-0 $QQ$ tetraquarks with lattice QCD data \cite{Bicudo:2024vxq,Lyu:2023xro}.
We treat the $0^{+}$ triplet meson mass $\Lambda_{0^{+}}$ as a free parameter to be fixed on the $T_{cc}^+\left(3875\right)$ mass. 

\paragraph{\textbf{Results for $T_{cc}^+\left(3875\right)$, $T_{bb}$, and $T_{bc}$.} }
The binding energy depends on the  $0^{+}$ triplet meson mass $\Lambda_{0^{+}}$ (isospin $I=0$). 
For $Q_1Q_2$, the ground-state multiplet includes the states $\{0^+, 1^+\}$ with $\{S=0, S=1\}$, 
which are degenerate without spin interactions. 
Solving the Schr\"odinger equation gives $\Lambda_{0^{+}}^* = 664$~MeV, 
resulting in a bound state about $323$~keV below the $DD$ threshold, 
which we identify with the $T_{cc}^+(3875)$
and a scattering length $a = 7.82$~fm, 
which is consistent with the LHCb result $a=7.16\pm 0.51$~fm~\cite{LHCb:2021auc}.
Notice that the operator that defines the triplet (see~\cite{Berwein:2024ztx}) coincides with a good diquark~\cite{Francis:2021vrr}.
Having fixed $\Lambda_{0^{+}}^*$, we predict a deeply bound $T_{bb}$ state ($J^P=1^+$), $116$~MeV  below the $BB$ threshold and a $T_{bc}$ state ($J^P=\{0^+, 1^+\}$) $24$~MeV below the $DB$ threshold.
Our $T_{bb}$ binding energy result aligns well with ab initio lattice studies \cite{Leskovec:2019ioa,Alexandrou:2024iwi,Aoki:2023nzp,Mohanta:2020eed,Junnarkar:2018twb,Francis:2016hui,Hudspith:2020tdf,Hudspith:2023loy,Colquhoun:2024jzh, Tripathy:2025vao} and lattice based BO approximation studies \cite{Brown:2012tm,Bicudo:2012qt,Bicudo:2015kna,Bicudo:2015vta,Bicudo:2016ooe}
Similarly, our $T_{bc}$ binding energy agrees well with lattice QCD predictions~\cite{Meinel:2022lzo, Francis:2018jyb, Padmanath:2023rdu, Radhakrishnan:2024ihu}.
A positive $30$~MeV spin splitting correction for $T_{bb}$ is reported in Ref. \cite{Bicudo:2016ooe}.

\paragraph{\textbf{Summary.}} 
The BOEFT predicts a set of coupled channel Schr\"odinger equations
to describe tetraquarks in QCD.
In particular, the $\chi_{c1}\left(3872\right)$ emerges as a bound state 
of coupled Schr\"odinger equations involving two tetraquark potentials and the quarkonium
potential, while the $T_{cc}^+\left(3875\right)$ emerges as a bound state of a single tetraquark potential.
The value of the adjoint mass and the triplet mass together with the structure of the potential, which are constrained by the BOEFT, originate states with a very  large radius, small binding energy 
and other properties compatible with experiments.
 The states are neither simple molecules nor compact tetraquarks but result from a conspiracy between the short- and long-range behavior of potentials--constrained by symmetry and by lattice QCD near the string-breaking region.
Thanks to the BOEFT factorization, fixing the parameters in the $cc$ sector  allows direct predictions in the $bb$ and $bc$ sectors. 
The existence of $T_{cc}^+\left(3875\right)$ below threshold may be related to the corresponding triplet meson being a good diquark \cite{Berwein:2024ztx}.

\begin{acknowledgments}
We acknowledge the DFG cluster of excellence ORIGINS funded by the Deutsche Forschungsgemeinschaft under Germany’s Excellence Strategy-EXC-2094-390783311.
N.~B. acknowledges the European Union ERC-2023-ADG-Project EFT-XYZ.  
We thank E. Braaten, T. Hatsuda, R. Mussa, and M. Scodeggio for their helpful discussions. 
We thank F. Knechtli, Y. Lyu, L. M\"uller for sharing their lattice data with us.
N.~B. acknowledges useful discussions during the long-term workshop, HHIQCD2024, at the Yukawa Institute for Theoretical Physics (YITP-T-24-02).
\end{acknowledgments} 

\bibliography{main}
\onecolumngrid
\begin{center}
\large\textbf{Supplemental Material for\\ \enquote{The nature of $\chi_{c1}\left(3872\right)$ and $T_{cc}^+\left(3875\right)$}}
\end{center}
\vspace{1cm}

\section{Schr\"odinger equation and multiplets for $\chi_{c1}\left(3872\right)$}\label{sec:SchX3872}

Here, we present the coupled Schr\"odinger equations relevant for the $\chi_{c1}\left(3872\right)$, which we identify with
 the $J^{PC}=1^{++}$ (isospin singlet $I=0$) state in the ground-state multiplet with $l=1$ and $S=1$, corresponding to the isospin $I=0$ $k^{PC} = 1^{--}$ adjoint meson (see Table \ref{tab:QQbarqqbar}). 
 We emphasize that the form of these  Schr\"odinger equations
 is a direct prediction of the BOEFT \cite{Berwein:2024ztx}.
 For  the definition of the adjoint meson mass and the adjoint meson mass  correlator as well as its appearence in the short distance multipole 
 expansion of the potentials see Sec. VB.1 of \cite{Berwein:2024ztx}.
 
 The coupled Schr\"odinger equations involve the  quarkonium BO potential with quantum number $\Sigma_g^+$, the tetraquark BO potentials with quantum numbers $\Sigma_g^{+\prime}$ and $\Pi_g$, which correspond to the $1^{--}$ adjoint meson at short distances and approach the $M\bar{M}$ (isospin singlet $\left(I=0\right)$) threshold at large distances, and the mixing potential $V_{\Sigma_{g}^{+}-\Sigma_{g}^{+\prime}}(r)$, which arises from the 
 mixing between the quarkonium $\Sigma_g^+$ and the tetraquark $\Sigma_g^{+\prime}$ BO potentials \cite{Berwein:2024ztx}. Our parameterization for these potentials are given by Eqs.~\eqref{eq:Vcornell}-\eqref{eq:Vmix}. The radial coupled Schr\"odinger equations\footnote{The potential matrix, excluding the third row and column relative to $V_{\Pi_g}$ in  Eq.~\eqref{eq:Sch-QQbar} has the same form as the model static Hamiltonian considered in Eq.~(6) of Ref.~\cite{Bulava:2024jpj}.} are given by \cite{Berwein:2024ztx}
\begin{align}
&\left[
-\frac{1}{m_Qr^2}\,\partial_rr^2\partial_r+\frac{1}{m_Qr^2}
{\begin{pmatrix}
l\left(l+1\right) & 0 & 0\\
0                 & l(l+1)+2        & -2\sqrt{l(l+1)} \\
0                 & -2\sqrt{l(l+1)} & l(l+1)
\end{pmatrix}}\right.
\nonumber\\
&\hspace{4.0 cm}\left.
+\begin{pmatrix} V_{\Sigma_{g}^{+}}(r) &  V_{\Sigma_{g}^{+}-\Sigma_{g}^{+\prime}}(r) & 0 \\
    V_{\Sigma_{g}^{+}-\Sigma_{g}^{+\prime}}(r) & V_{\Sigma_{g}^{+\prime}}(r) & 0\\
      0 & 0 & V_{\Pi_g}(r)\end{pmatrix}
      \right]
      \hspace{-4pt}\begin{pmatrix} \psi_{\Sigma} \\ \psi_{\Sigma^{\prime}} \\ \psi_{\Pi }\end{pmatrix}={\mathcal{E}} \begin{pmatrix} \psi_{\Sigma} \\ \psi_{\Sigma^{\prime}} \\ \psi_{\Pi}\end{pmatrix},
      \label{eq:Sch-QQbar}
\end{align}
where ${\mathcal{E}}$ is the eigenenergy. Eq.~\eqref{eq:Sch-QQbar} contains both the short distance mixing between static energies with quantum numbers $\Sigma_g^{+\prime}$ and $\Pi_g$ which leads to off-diagonal terms in the kinetic matrix, corresponding to the $1^{--}$ adjoint meson and the long distance mixing  between
energies with the same BO quantum numbers $\Sigma_g^+$, which leads to off-diagonal terms in the diabatic potential matrix (see Ref.~\cite{Berwein:2024ztx} for details).
They are two different effects that dominate in different distance regions. For quarkonium, the angular momentum is $l=l_Q$, where $l_Q\left(l_Q+1\right)$ is the eigenvalue of the heavy quark pair orbital angular momentum ${\bm L}_Q^2$.
The  corresponding values of the angular momentum $l$ for the $Q\bar{Q}q\bar{q}$ states  are given in the fourth column of Table~\ref{tab:QQbarqqbar}.

The adiabatic static energies with quantum numbers $1\Sigma_g^+$, $2\Sigma_g^+$, and $1\Pi_g$ shown in Fig.~\ref{fig:isospin0} are the eigenvalues of the diabatic potential matrix in Eq.~\eqref{eq:Sch-QQbar} and  are given by
\begin{align}
 V_{1\Sigma_g^+ }(r) &= \frac{V_{\Sigma_g^+}\left(r\right)+ V_{\Sigma_g^{+\prime}}\left(r\right)} {2}-\sqrt{\left(\frac{V_{\Sigma_g^+}\left(r\right)- V_{\Sigma_g^{+\prime}}\left(r\right)}{2}\right)^2+\,V^2_{{\scaleto{\Sigma_g^+-\Sigma_g^{+\prime}\mathstrut}{6pt}}}(r)},\label{eq:V1S}\\
V_{2\Sigma_g^+ }(r) &= \frac{V_{\Sigma_g^+}\left(r\right)+ V_{\Sigma_g^{+\prime}}\left(r\right)} {2}+\sqrt{\left(\frac{V_{\Sigma_g^+}\left(r\right)- V_{\Sigma_g^{+\prime}}\left(r\right)}{2}\right)^2+\,V^2_{{\scaleto{\Sigma_g^+-\Sigma_g^{+\prime}\mathstrut}{6pt}}}(r)},
    \label{eq:V2S}\\
V_{1\Pi_g}(r) &= V_{\Pi_g}\left(r\right).
   \label{eq:V1P}
\end{align}
More details on the diabatic to adiabatic transformation can be found in appendix~A of Ref.~\cite{Berwein:2024ztx}.

\begin{table}[h!]
\centering
\resizebox{0.70\columnwidth}{!}{%
	\begin{tabular}{||c|c|c||c|c|c||}
		\hline\hline
		\multirow{2}{*}{\hspace{2pt}$\begin{array}{c} Q\bar{Q}\\\text{color state}\end{array}$\hspace{2pt}} & \multirow{2}{*}{\hspace{2pt}$\begin{array}{c}\text{$q\bar{q}$ spin}\\k^{PC}\end{array}$\hspace{2pt}} & \multirow{2}{*}{\hspace{2pt} $\begin{array}{c} \text{BO quantum \#}\\\Lambda^\sigma_\eta\end{array}$\hspace{2pt}}&\multirow{2}{*}{\hspace{2pt} $l$\hspace{2pt}}& \multirow{2}{*}{\hspace{2pt} $\begin{array}{c}J^{PC}\\\{S=0, S=1\}\end{array}$\hspace{2pt}}& \multirow{2}{*}{\hspace{2pt}Multiplets\hspace{2pt}}\\
		& & & & &  \\
		\hline\hline
		\multirow{6}{*} {\hspace{2pt}$\begin{array}{c} \text{Octet}\\\mathbf{8}\end{array}$\hspace{2pt}}
		&\multirow{3}{*}{$0^{-+}$} & \multirow{3}{*}{$\Sigma_u^-$}  & \hspace{2pt}$0$\hspace{2pt} & \hspace{2pt}$\{0^{++}, 1^{+-}\}$\hspace{2pt}
		&\hspace{2pt}$T_1^0$\hspace{2pt}\\
		\cline{4-6}
		& & & \hspace{2pt}$1$\hspace{2pt} & \hspace{2pt}$\{1^{--}, \left(0, 1, 2\right)^{-+}\}$\hspace{2pt}&\hspace{2pt}$T_2^0$\hspace{2pt}\\
		\cline{2-5}\cline{2-6}
		&\multirow{4}{*}{$1^{--}$} & ${\Sigma_g^{+\prime},\Pi_g}$  & \hspace{2pt}$1$\hspace{2pt} & \hspace{2pt}$\{1^{+-}, (0,1,2)^{++}\}$\hspace{2pt}&\hspace{2pt}$T_1^1$\hspace{2pt} \\
		\cline{3-6}
		& & ${\Sigma_g^{+\prime}}$  & \hspace{2pt}$0$\hspace{2pt} & \hspace{2pt}$\{0^{-+},  1^{--}\}$\hspace{2pt}&\hspace{2pt}$T_2^1$\hspace{2pt} \\
		\cline{3-6}
		& & ${\Pi_g}$  & \hspace{2pt}$1$\hspace{2pt} & \hspace{2pt}$\{1^{-+},  (0,1,2)^{--}\}$\hspace{2pt}&\hspace{2pt}$T_3^1$\hspace{2pt} \\
		\cline{3-6}
		\hline\hline
	\end{tabular}}
	\caption{$J^{PC}$ multiplets for the lowest $Q\bar{Q}q\bar{q}$ tetraquarks \cite{Berwein:2024ztx}. The third
		column shows the  BO quantum numbers corresponding to the potentials that appear in the Schr\"odinger equation for the respective
		multiplet.  We label the multiplet as $T_i^k$ in the last column in the order of increasing energies, where $k(k+1)$ is the eigenvalue corresponding to the LDF total angular momentum ${\bm K}^2$. In absence of spin interactions, different spin combinations $\{S=0, S=1\}$
        form degenerate multiplets. We identify $\chi_{c1}(3872)$ state with $J^{PC}=1^{++}$  state in the ground-state $T^1_1$ multiplet corresponding to $k^{PC} = 1^{--}$.}
	\label{tab:QQbarqqbar}
\end{table}

Previous studies on $\chi_{c1}\left(3872\right)$ within the BO approximation has been done in Refs.~\cite{Braaten:2024tbm, Bruschini:2020voj, Bruschini:2022bsh}. In \cite{Braaten:2024tbm}, the $\chi_{c1}\left(3872\right)$ state is considered as a bound state in the pair of BO potentials with quantum numbers $\Sigma_g^{+\prime}$ and $\Pi_g$ associated with the  isospin $I=0$ $k^{PC} = 1^{--}$ adjoint meson but the narrow avoided crossing with the quarkonium $\Sigma_g^+$ BO potential was ignored and the spin splittings in the tetraquark multiplet ($T^1_1$ in Table~\ref{tab:QQbarqqbar}) was included by treating the spin-splittings in charm meson pair states within first-order perturbation theory (see Eq.~\eqref{eq:VSS}). In Refs.~\cite{Bruschini:2020voj, Bruschini:2022bsh}, the $\chi_{c1}\left(3872\right)$ state is obtained as a bound state from a coupled channel Schr\"odinger equation where the heavy meson pair threshold $M\bar{M}$ was treated as a constant energy for all values of $r$, where $r$ is the separation between heavy quark and antiquark pairs, instead of being associated with the $1^{--}$ adjoint meson at short-distance $\left(r\rightarrow 0\right)$, which was only realized recently \cite{Berwein:2024ztx}. The avoided crossing between the heavy meson pair threshold $M\bar{M}$ and the quarkonium BO potential $V_{\Sigma_g^+}$ was included but the mixing potential $V_{\Sigma_{g}^{+}-\Sigma_{g}^{+\prime}}(r)$ was parameterized by a gaussian potential whose parameters were fixed to get a $\chi_{c1}\left(3872\right)$ state just below $D^{*0}\bar{D}^0$ threshold.  Moreover, a detailed analysis of the scattering amplitude up to center of mass energy $4$~GeV in \cite{Bruschini:2022bsh} predicted an unobserved conventional charmonium resonance $\chi_{c1}\left(2P\right)$ with $J^{PC}=1^{++}$ and mass around $3960$~MeV in addition to the $\chi_{c1}\left(3872\right)$. Additionally, studies on $\chi_{c1}\left(3872\right)$ within the BO approximation inspired framework has been done in Refs.~\cite{Lebed:2022vks, Lebed:2023kbm} where the state was considered to be in the ground state multiplet of the diquark $\left(\bar{Q}\bar{q}\right)_{3}$ and antidiquark $\left(Qq\right)_{\bar{3}}$ configuration.
Our work is the first full application of the BOEFT prediction 
for the coupled Schr\"odinger equations and the form of the potentials
at short and long range, which includes also the lattice input 
(see Fig.~\ref{fig:isospin0}), existing for the moment only in the intermediate 
distance region.

\section{Schr\"odinger equation and multiplets for $T_{cc}^+\left(3875\right)$}\label{sec:SchTcc3875}

Here, we present the Schr\"odinger equation relevant for the $T_{cc}^+\left(3875\right)$, which we identify with $J^{PC}=1^{+}$  state in the ground-state multiplet with $l=0$ and $S=1$, corresponding to $k^{P} = 0^+$ triplet meson (see Table \ref{tab:QQqq}). The Schr\"odinger equation involve the BO potential with quantum number $\Sigma_g^+$ which correspond to the $0^{+}$ triplet meson at short distances and approach the $MM$ (isospin singlet $\left(I=0\right)$) threshold at large distances \cite{Berwein:2024ztx}. 
We emphasize that it is a prediction of the BOEFT 
that in this case  only one radial
 Schr\"odinger contributes \cite{Berwein:2024ztx}.
 For  the definition of the triplet meson mass and the triplet meson mass  correlator as well as its appearence in the short distance multipole   expansion of the potentials see Sec. VB.2 of \cite{Berwein:2024ztx}.
The
Schr\"odinger equation is given by \cite{Berwein:2024ztx}
\begin{align}
	\left[-\frac{1}{m_Qr^2}\,\partial_r\,r^2\,\partial_r+\frac{l(l+1)}{m_Qr^2}+V_{\Sigma_{g}^+}(r)\right]\psi_{\Sigma_g^+}=\mathcal{E}\,\psi_{\Sigma_g^+}\,,
	\label{eq:SchQQ}
\end{align}
${\mathcal{E}}$ is the eigenenergy and our parameterization for the BO potential $V_{\Sigma_g^+}(r)$ is given by Eq.~\eqref{eq:VQQ}.

\begin{table}[h!]
\centering
\resizebox{0.70\columnwidth}{!}{%
	\begin{tabular}{||c|c|c|c|c||c|c||}
		\hline\hline
		\multirow{2}{*}{\hspace{2pt}$\begin{array}{c} QQ\\\text{ color state}\end{array}$\hspace{2pt}} & \multirow{2}{*}{\hspace{2pt}$\begin{array}{c} \text{$\bar{q}\bar{q}$ spin}\\ k^{P}\end{array}$\hspace{2pt}} & \multirow{2}{*}{\hspace{2pt} $\begin{array}{c} \text{BO quantum \#}\\\Lambda^\sigma_\eta\end{array}$\hspace{2pt}} & \multirow{2}{*}{\hspace{2pt}$\begin{array}{c} \text{Isospin}\\I\end{array}$ \hspace{2pt}}& \multirow{2}{*}{\hspace{2pt} $l$\hspace{2pt}} & \multicolumn{2}{|c||}{$J^{P}$}\\
		\cline{6-7}
		& & & & & \hspace{2pt}$S=0$\hspace{2pt} & \hspace{2pt}$S=1$\hspace{2pt} \\
		\hline\hline
		\multirow{4}{*}{\hspace{2pt}$\begin{array}{c} \text{Antitriplet}\\\bar{\mathbf{3}}\end{array}$\hspace{2pt}} &\multirow{2}{*}{$0^+$} & \multirow{2}{*}{${\Sigma_g^+}$} & \multirow{2}{*}{$0$}& $0$ & \hspace{2pt} --- & \hspace{2pt}$1^+$\hspace{2pt} \\
		\cline{5-7}
		& & & & 1 &\hspace{2pt}$1^-$\hspace{2pt}&---\\
		\cline{2-7}
		& \multirow{2}{*}{$1^+$} & \multirow{2}{*}{${\Sigma_g^-, \Pi_g}$} & \multirow{2}{*}{$1$}& $0$ &  \hspace{2pt}$0^-$\hspace{2pt}& --- \\
		\cline{5-7}
		& & & & 1 & $1^{-}$&\hspace{2pt}$\left(0, 1, 2\right)^+$\hspace{2pt}\\
		\cline{5-7}
		\hline\hline
	\end{tabular}}
	\caption{$J^{PC}$ multiplets for the lowest $QQ\bar{q}\bar{q}$ tetraquarks \cite{Berwein:2024ztx} considering only the color antitriplet configuration $\left(QQ\right)_{\bar{3}}$. Since, the $\left(QQ\right)_{\bar{3}}$ color antitriplet potential is attractive while $\left(QQ\right)_{6}$ color sextet potential is repulsive, we expect the states with $\left(QQ\right)_{\bar{3}}$ configuration to be lower in energy.  The third
    column shows the  BO quantum numbers corresponding to the potentials that appear in the Schr\"odinger equation for the respective multiplet.
    In the last column, the dashed entry means that that particular state is not allowed due to the Pauli exclusion principle. We identify $T_{cc}^+\left(3875\right)$ state with $J^{PC}=1^{+}$  state in the ground-state multiplet with $l=0$ and $S=1$, corresponding to $k^{P} = 0^+$. }
	\label{tab:QQqq}
\end{table}

Previous studies on $T_{cc}^+\left(3875\right)$ within the BO approximation inspired framework has been done in Ref.~\cite{Lebed:2024zrp} where the state was considered to be in the ground state multiplet of the diquark $\Delta\equiv \left(\bar{q}\bar{q}\right)_{3}$ and antidiquark $\bar{\Delta}\equiv\left(QQ\right)_{\bar{3}}$ configuration.  The authors solve a multi-channel Schr\"odinger equation accounting for the spin and isospin splittings in the $DD$ thresholds. For the multi-channel Schr\"odinger equation, they considered a potential matrix where the diagonal elements included the Cornell type (perturbative Coulomb plus linear potential) potential between the $\Delta-\bar{\Delta}$ and the sum of the masses of the thresholds $D^0D^{*+}$ and $D^+D^{*0}$, while the off-diagonal elements are the mixing potentials to account for the mixing between the  $\Delta-\bar{\Delta}$ and $DD$ thresholds due to string breaking. Note that, in BOEFT, we also consider the diquark $\Delta\equiv \left(\bar{q}\bar{q}\right)_{3}$ and antidiquark $\bar{\Delta}\equiv\left(QQ\right)_{\bar{3}}$ configuration as shown in Table~\ref{tab:QQqq},  however, BOEFT gives a single channel Schr\"odinger Eq.~\eqref{eq:SchQQ} that involves the BO potential with quantum $\Sigma_g^+$. The BOEFT predicts that based on the BO quantum number conservation, the potential $V_{\Sigma_g^+}(r)$ has an attractive behavior at short-distance due to color antriplet $\left(QQ\right)_{\bar{3}}$ shifted by the $0^+$ triplet meson mass ($\Lambda_{0^+}$ in Eq.\eqref{eq:VQQ}) and smoothly evolve to the $DD$ threshold at large distances as shown in Fig.~\ref{fig:isospin1}. Unlike \cite{Lebed:2024zrp}, there is no avoided crossing between the $V_{\Sigma_g^+}(r)$ and the $DD$ threshold due to string breaking, which is also consistent with the lattice results shown in Fig.~\ref{fig:isospin1}.

Previous studies on $T_{bb}$ state with $J^P=1^+$ within the BO approximation have been performed in  Refs.~\cite{Brown:2012tm,Bicudo:2012qt,Bicudo:2015kna,Bicudo:2015vta,Bicudo:2016ooe}, where  the BO potential $V_{\Sigma_g^+}(r)$ between the heavy quarks was computed as a function of separation $r$ using lattice QCD. In Ref.~\cite{Brown:2012tm}, the BO potential $V_{\Sigma_g^+}(r)$ was parameterized with a quark model potential form for the two heavy-light meson system, which included the perturbative Coulomb, spin-spin, and linear confinement terms. The Coulomb potential had an attractive $1/r$ form smeared with a Gaussian.
Additionally, a simple Yukawa-like term accounting for one-pion exchange with pion mass around $390$~MeV was also included to account for long-range meson exchange interactions. With this potential, the authors in \cite{Brown:2012tm} solved a single-channel Schr\"odinger equation (Eq.~\eqref{eq:SchQQ}) and found a single bound state with $50$~MeV binding energy.  In Refs.~\cite{Bicudo:2012qt,Bicudo:2015kna,Bicudo:2015vta}, the BO potential $V_{\Sigma_g^+}(r)$ was parametrized by an attractive Coulomb potential $1/r$ smeared with a Gaussian plus a constant that  corresponds to twice the mass of static heavy-light meson. Even though this  potential is consistent with the BO quantum number $\Lambda^{\sigma}_{\eta}$ conservation constraints in the $r\rightarrow0$  and $r\rightarrow\infty$ limits, it is inconsistent with the pNRQCD short-distance $\left(r\rightarrow0\right)$ multipole expansion compared to our parameterization in Eq.~\eqref{eq:VQQ}. With this potential, the authors in \cite{Bicudo:2012qt,Bicudo:2015kna,Bicudo:2015vta} solved a single-channel Schr\"odinger equation (Eq.~\eqref{eq:SchQQ}) and found a $T_{bb}$ state around $95$~MeV below the spin-isospin averaged $BB$ threshold, which agrees with our prediction considering the error bars in Refs.~\cite{Bicudo:2012qt,Bicudo:2015kna,Bicudo:2015vta}. 

\section{Spin-splitting}\label{sec:Spin-QQbar}

At leading order in $1/m_Q$ expansion, the BOEFT-potential includes only static contributions, and solving the Schr\"odinger equation in this potential yields a degenerate (spin-averaged) set of states independent of the heavy quark spin ${\bm S}$. However, the BOEFT allows  also to calculate the form of the spin-dependent corrections 
to the potential. Differently from quarkonium, in hybrids and in tetraquarks, spin-dependent corrections
already arise at order $1/m_Q$  \cite{Oncala:2017hop, Brambilla:2018pyn, 
Brambilla:2019jfi, Soto:2023lbh,Schlosser:2025tca,Soto:2020xpm}. 
The general form of the spin-corrections to the BO potentials in BOEFT 
for the tetraquarks and the hybrids
is known \cite{Soto:2023lbh},
but involves a lattice calculation of generalized static Wilson 
loops. At the moment only preliminary lattice results
for these quantities for the  case of hybrids  exist\cite{Schlosser:2025tca}.
In \cite{Brambilla:2018pyn, Brambilla:2019jfi}, the whole set of hybrid spin-dependent potentials was obtained in a two steps procedure, incorporating both the heavy quark flavor-dependent perturbative contributions and the flavor-independent LDF nonperturbative contributions, up to order $1/m_Q^2$. Using these potentials, the spin-splitting in hybrid multiplets was determined. The nonpertubative contributions in the spin-dependent potentials were extracted from the direct lattice calculations of  charmonium hybrid multiplets \cite{HadronSpectrum:2012gic, Cheung:2016bym}  and subsequently used to predict bottomonium hybrid spin multiplets, which are well in agreement with the recent lattice QCD results for bottomonium hybrid states \cite{Ryan:2020iog}.

Note that, for adjoint meson $k^{PC}=1^{--}$,  the ground-state $Q\bar{Q}$ tetraquark multiplet has $l=1$ ($T_1^1$ multiplet in Table~\ref{tab:QQbarqqbar}) and  includes a $S=0$ state $1^{+-}$ and three  $S=1$ states $(0,1,2)^{++}$ (see Table~\ref{tab:QQbarqqbar}): $\chi_{c1}\left(3872\right)$ is identified with $1^{++}$ state. In the current work, due to absence of the direct lattice QCD calculations for the spin-splittings of the tetraquark multiplet, we estimated the spin corrections for the lowest $c\bar{c}$ tetraquark by averaging the lattice results from the two references \cite{HadronSpectrum:2012gic, Cheung:2016bym} for the spin-splittings of the lowest $c\bar{c}$ hybrid multiplet (see Fig.~\ref{fig:spin1--}). Similarly,  the lattice QCD calculations for the spin-splittings of the lowest $b\bar{b}$ hybrid multiplet in \cite{Ryan:2020iog} can be used for estimating the spin-splittings for the lowest $b\bar{b}$ tetraquark multiplet.

Alternatively, one could estimate the spin splittings within ground-state $Q\bar{Q}$ tetraquark multiplet in the BO approximation by exploiting the spin splittings in the s-wave heavy-light meson  pair states $M\bar{M}$ \footnote{For this work, $M\bar{M}$ refers to $D\bar{D}$ threshold in the charm sector and $B\bar{B}$ in the bottom sector} \cite{Braaten:2024tbm}. In fact, the BO potentials for the $Q\bar{Q}$ tetraquark at large $r$ (where $r$ is the separation between $Q$ and $\bar{Q}$), evolve into the heavy-light meson pair $M\bar{M}$ energy in the static limit due to BO quantum number conservation \cite{Berwein:2024ztx}. The spin-dependent interactions that leads to spin-splitting in the $M\bar{M}$ threshold also appears at order $1/m_Q$. Exploiting  this  fact,
the order $1/m_Q$ spin-dependent  BO potential for tetraquark simplifies to a potential  $V_{SS}$ that splits the $M\bar{M}$ thresholds \cite{Braaten:2024tbm}: 
\begin{equation}
 V_{SS} = \delta_{Q}\left({\bm S}_1.{\bm K}_1+{\bm S}_2.{\bm K}_2\right)
 \label{eq:VSS}
\end{equation}
where $\bm{S}_1$ and ${\bm K}_1$ denotes the heavy quark spin and light antiquark angular momentum in meson $M$, $\bm{S}_2$ and ${\bm K}_2$ denotes the heavy antiquark spin and light quark angular momentum in meson $\bar{M}$, and $\delta_Q$ represents the spin splitting between the heavy mesons $M^*$ and $M$: $\delta_c=141$~MeV and $\delta_b=45$~MeV~\cite{ParticleDataGroup:2024cfk}. Given the spin-dependent potential $V_{SS}$ in Eq.~\eqref{eq:VSS}, the  thresholds for s-wave heavy-light meson pair states   are $-\frac{3}{2} \delta_Q$ for $M \bar M$, $-\frac{1}{2} \delta_Q$ for $M^*\bar M$ and $M \bar M^*$, and $+\frac{1}{2} \delta$ for $M^*\bar M^*$.

A first application using Eq.~\eqref{eq:VSS}  for estimating the  spin splittings in the ground-state  $Q\bar{Q}$ tetraquark multiplet was done in Ref.~\cite{Braaten:2024tbm}.
Using Fierz transformations, the state in the ground-state multiplet $\{1^{+-}, \left(0, 1, 2\right)^{++}\}$ (see Table~\ref{tab:QQbarqqbar}) can be expressed as a superposition of s-wave heavy-light meson pair states $M\bar{M}$, $M^*\bar{M}$ or $M\bar{M}^*$, and $M^*\bar{M}^*$. Treating $V_{SS}$ within first-order perturbation theory,  
the authors in Ref.~\cite{Braaten:2024tbm} finds that the spin-singlet ($S=0$) $1^{+-}$ state receives no corrections at first order in $\delta_Q$ and lies halfway  between $M^*\bar{M}$ and $M^*\bar{M}^*$ thresholds while the spin-triplet states ($S=1$) states $(0,1,2)^{++}$: $0^{++}$ shifted by $-\delta_Q$, $1^{++}$ shifted by $-\delta_Q/2$ and $2^{++}$ shifted by $+\delta_Q/2$ with respect to the spin-averaged threshold. This implies that $0^{++}$ lies halfway  between $M\bar{M}$ and $M^*\bar{M}$ thresholds  while the $1^{++}$ and $2^{++}$ states remain near the $M^*\bar{M}$ and  $M^*\bar{M}^*$ threshold respectively, which is well in agreement with our prediction for $c\bar{c}$ tetraquark shown  Fig.~\ref{fig:spin1--}. 

With respect to $c\bar{c}$ tetraquark states, 
we found a spin-averaged bound state about $90$~keV below spin-isospin averaged $D\bar{D}$ threshold for the  adjoint meson mass $\Lambda^*_{1^{--}}=919$~MeV after solving the coupled channel Schr\"odinger equation in Eq.~\eqref{eq:Sch-QQbar} with the potential parameterizations in Eqs.~\eqref{eq:Vcornell}-\eqref{eq:Vmix}. Using the spin-splitting results from Ref.~\cite{Braaten:2024tbm}, the $1^{+-}$ state will be at the spin-isospin averaged $D\bar{D}$ threshold which agrees with our prediction of $11(11)$ MeV within error bars, while the  $1^{++}$ and $2^{++}$ states will be shifted below by  $70.5$~MeV and shifted above by $70.5$~MeV respectively relative to the spin-isospin averaged $D\bar{D}$ threshold, which again agrees with our predictions of $74(14)$~MeV below for $1^{++}$  and $58(14)$~MeV above for $2^{++}$ within error bars. The $0^{++}$ state is shifted below by $141$~MeV relative to spin-isospin averaged threshold which is consistent with our prediction of $100(11)$~MeV below within error bars. 

With respect to $b\bar{b}$ tetraquark states, we found a spin-averaged bound state about $15$~MeV below spin-isospin averaged $B\bar{B}$ threshold for the adjoint meson mass fixed to $\Lambda^*_{1^{--}}=919$~MeV after solving the coupled channel Schr\"odinger equation in Eq.~\eqref{eq:Sch-QQbar}. Again using the lattice results for the spin-splittings of the lowest $b\bar{b}$ hybrid multiplet from Ref.~\cite{Ryan:2020iog}, the  $1^{+-}$ and $2^{++}$ states are shifted above by $0\left(22\right)$~MeV and $15\left(27\right)$~MeV, 
while the $0^{++}$ and $1^{++}$ states are shifted below by $26\left(16\right)$~MeV and $17\left(16\right)$~MeV 
relative to the spin averaged. The central value of $\chi_{b1}\left(1^{++}\right)$ state is around $10$~MeV below the $B\bar{B}^*$ threshold after including the spin splitting but including the lattice error bar, we cannot exclude the shallow bound state. Using the spin-splitting results from Ref.~\cite{Braaten:2024tbm}, the $1^{+-}$ state will receive no corrections at first order in $\delta_Q$, while the $0^{++}$ and $1^{++}$ states will be shifted below by $45$~MeV and $22.5$~MeV relative to the spin-isospin averaged threshold, and the $2^{++}$ state will be shifted above by $22.5$~MeV related to the spin averaged. Again, we find that the spin-splitting estimates for the $b\bar{b}$ tetraquark states using the lattice spin-splittings results for the lowest $b\bar{b}$ hybrids and from Ref.~\cite{Braaten:2024tbm} are consistent within error bars. 

With respect to $T_{cc}^+\left(3875\right)$ and $T_{bb}$ states with $J^P=1^+$, we found a  bound state about $323$~keV and $116$~MeV below spin-isospin averaged $DD$ and  $BB$ threshold respectively  for the triplet meson mass fixed to $\Lambda^*_{0^{+}}=664$~MeV after solving the single channel Schr\"odinger equation in Eq.~\eqref{eq:SchQQ} with the potential parameterizations in Eq.~\eqref{eq:VQQ}. Using Eq.~\eqref{eq:VSS}, which splits the spin-isospin averaged $DD$ and $BB$ thresholds into three thresholds $MM$, $MM^*$ or $M^*M$, and $M^*M^*$ where $M=\left(D,B\right)$, we find that the $1^+$ state receives no corrections at first order in $\delta_Q$ unlike in Ref.~\cite{Bicudo:2016ooe}, where around positive $30$~MeV correction to $T_{bb}$ binding energy was reported by solving a coupled-channel Schr\"odinger equation accounting non-perturbatively for the effects of the bottom heavy-light  meson pair splittings. With respect to $T_{bc}$ states with $J^P=\{0^+, 1^+$\}, we found a  bound state about $24$~MeV below spin-isospin averaged $DB$ threshold for $\Lambda^*_{0^{+}}=664$~MeV. Using Eq.~\eqref{eq:VSS}, we find that both the states receive no spin corrections within first-order perturbation theory in $\delta_Q$.


\end{document}